\documentclass[prl,showpacs,superscriptaddress,twocolumn]{revtex4}
\usepackage{amsfonts}
\usepackage{amsmath}
\usepackage{amsthm}
\usepackage{amscd}
\usepackage{amssymb}
\usepackage{subfigure}
\usepackage{amsxtra}
\usepackage{bm}           % bold math
\usepackage{bbm}
\usepackage{graphicx}
\usepackage{epstopdf}
\usepackage{color}

\def\bi#1\ei {\begin{itemize}#1\end{itemize}}
\def\bn#1\en {\begin{enumerate}#1\end{enumerate}}
\def\bea#1\eea {\begin{align}#1\end{align}}
\def\bean#1\eean {\begin{align*}#1\end{align*}}
\def\ben#1\een {\begin{equation*}#1\end{equation*}}
\def\be#1\ee {\begin{equation}#1\end{equation}}
\def\bes#1\ees {\begin{equation}\begin{split}#1\end{split}\end{equation}}
\def\bear#1\eear {\begin{eqnarray}#1\end{eqnarray}}
\def\bear#1\eear {\begin{eqnarray*}#1\end{eqnarray*}}

\newcommand{\beq}{\begin{equation}}
\newcommand{\eeq}{\end{equation}}

\newcommand{\ket}[1]{\ensuremath{\left|#1\right\rangle}}
\newcommand{\braket}[2]{\ensuremath{\langle #1|#2\rangle}}
\newcommand{\ketbra}[2]{\ensuremath{| #1 \rangle \langle #2 |}}

\newcommand{\Hil}{\mathcal{H}}

\newcommand{\id}{\ensuremath{\mathbbm 1}}

\renewcommand{\qed}{\ensuremath{\hfill \blacksquare}}
\newcommand{\order}[1]{\mathcal{O}(#1)}
\newcommand{\am}{\ensuremath{\tfrac{\alpha}{\sqrt{m}}}}
\newcommand{\amsqr}{\ensuremath{\tfrac{|\alpha|^2}{m}}}
\newtheorem{thm}{Theorem}

\begin{document}

\title{Quantum Fingerprinting with Coherent States and a\\ Constant Mean Number of Photons}

\author{Juan Miguel Arrazola}
\author{Norbert L\"utkenhaus}

\affiliation{Institute for Quantum Computing and Department of Physics and Astronomy,
University of Waterloo, 200 University Avenue West,
N2L 3G1 Waterloo, Ontario, Canada}

\begin{abstract}
We present a protocol for quantum fingerprinting that is ready to be implemented with current technology and is robust to experimental errors. The basis of our scheme is an implementation of the signal states in terms of a coherent state in a superposition of time-bin modes. Experimentally, this requires only the ability to prepare coherent states of low amplitude, and to interfere them in a balanced beam splitter. The states used in the protocol are arbitrarily close in trace distance to states of $\order{\log_2 n}$ qubits, thus exhibiting an exponential separation in abstract communication complexity compared to the classical case. The protocol uses a number of optical modes that is proportional to the size $n$ of the input bit-strings, but a total mean photon number that is constant and independent of $n$. Given the expended resources, our protocol achieves a task that is provably impossible using classical communication only.  In fact, even in the presence of realistic experimental errors and loss, we show that there exist a large range of input sizes for which our quantum protocol transmits an amount of information that can be more than two orders of magnitude smaller than a classical fingerprinting protocol.
\end{abstract}

\date{\today}

\maketitle
\textit{Introduction.-} Communication complexity is the study of the amount of communication that is required to perform distributed information-processing tasks. This corresponds to the scenario in which two parties, Alice and Bob, respectively receive inputs $x,x'\in\{0,1\}^n$. Their goal is to collaboratively compute the value of a boolean function $f(x,x')$ with as little communication as possible \cite{Yao1979}. Although they can always do this by communicating their entire input, there are many situations in which they can succeed with significantly less communication \cite{kushilevitz2006communication}. 

Likewise, quantum communication complexity studies the case where the parties are allowed to employ quantum resources such as quantum channels and shared entanglement (see Refs. \citep{BrassardQCC,RevModPhys.82.665} for an overview). Remarkably, it has been proven that there exist various problems for which the use of quantum resources offer exponential savings in communication compared to their classical counterparts \cite{DJexp,RazProblem,HM-Bar-Yossef,HM-Gavinsky,QuantumFingerprinting}. Unfortunately, these results are currently accompanied by only a few experimental demonstrations \cite{ZhangQCC,TrojekQCC,HornFPs}, and providing a method to facilitate their implementation is a pressing problem.

We focus on the \textit{simultaneous message passing} model \cite{Yao1979}, in which Alice and Bob are not allowed to communicate with each other but instead send messages to a third party, the referee, who must determine the value of the function based only on the messages she receives. An important example is the equality problem, where $f(x,x')=1$ if and only if $x=x'$. In this case, Alice and Bob can achieve their goal by sending much shorter \textit{fingerprints} of their original inputs. If they are restricted to classical messages and local randomness only, it has been shown that the optimal classical protocols require fingerprints of length at least $\Omega(\sqrt{n})$ when an arbitrarily small probability of error is allowed \cite{ambainis1996communication,babai1997randomized,newman1996public}. On the other hand, it was shown in Ref. \cite{QuantumFingerprinting} that if Alice and Bob are allowed to send quantum states, then they only need to send fingerprints of $\order{\log n}$ qubits, thus demonstrating an exponential separation between classical and quantum communication complexity.

In this work, we present a protocol for quantum fingerprinting that uses quantum states that are arbitrarily close in trace distance with respect to states of $\order{\log_2 n}$ qubits, thus exhibiting an exponential separation in abstract communication complexity compared to the classical case. The protocol is robust to experimental imperfections and is characterized by a probability of error which is tunable and can be made arbitrarily small. Moreover, in an ideal implementation, the mean photon number of the signals is independent of $n$, so that the energy cost of the protocol is constant regardless of the size of the messages.

In the remainder of this paper, we describe the results of Ref. \cite{QuantumFingerprinting} and, based on them, we outline the protocol for implementing quantum fingerprinting with coherent states and a constant mean number of photons. We then show how the protocol can be adjusted to account for experimental errors and we analyze its performance in realistic scenarios. Finally, we conclude by discussing further possible applications of our results as well as some of its limitations. 

\textit{Coherent-state quantum protocol.-} Quantum fingerprinting, as introduced in Ref. \cite{QuantumFingerprinting}, relies on the concept of error-correcting codes. A code can be expressed as a function $E:\{0,1\}^n\rightarrow\{0,1\}^m$, where $E(x)$ is the \textit{codeword} associated with the input $x$, and $m=cn$ for some $c>1$. The protocol makes use of codes that have the additional property that the minimum Hamming distance between any two codewords is at least $(1-\delta)m$, for some $\delta>0$. One example are Justesen codes \cite{justesen1972class}, for which we can have $\delta<\tfrac{9}{10}+\tfrac{1}{15c}$ whenever $c>2$. In Ref. \cite{QuantumFingerprinting}, a protocol is specified in which, for each possible input $x$ and corresponding codeword $E(x)$, Alice and Bob prepare the fingerprint states
\beq\label{FPstate}
\ket{h_x}=\frac{1}{\sqrt{m}}\sum_{i=1}^m(-1)^{E(x)_i}\ket{i},
\eeq
where $E(x)_i$ is the $i$th bit of the codeword $E(x)$. This state has dimension $m$, so it can be associated to a system of $\log_2 m=\order{\log_2 n}$ qubits.

An approach to implementing the fingerprint states is to decompose the underlying Hilbert space as a tensor product of Hilbert spaces of smaller dimension \cite{HornFPs,NMRFingerprinting,1qubitfp}. For example, we could have a collection of $\order{\log_2 n}$ two-level systems, such as photons in the polarization degree of freedom. As noted already in Ref. \cite{HornFPs}, a serious drawback of this strategy is that most fingerprint states must be highly entangled \cite{MoraKolmogorov,AlgoComplexity}, so that even for low input sizes, the experimental requirements greatly exceed that which is possible to achieve with current technology, except for the case of single-qubit quantum fingerprinting \cite{NMRFingerprinting,1qubitfp}.

Alternatively, we can consider the underlying Hilbert space as arising directly from a single $m-$dimensional physical system, such as a single photon distributed over $m$ orthogonal optical modes, as has been considered in Refs. \cite{QFingerprintingMassar,SWAPHOM}. In that case, let $b_i$ be the annihilation operator of the $i$th optical mode. We define the \textit{fingerprint mode} as $a_x=\frac{1}{\sqrt{m}}\sum_{i=1}^m(-1)^{E(x)_i}b_i$, so that a single-photon state in the fingerprint mode 
\beq\label{FPFockstate}
a^{\dagger}_x\ket{0}=\frac{1}{\sqrt{m}}\sum_{i=1}^m(-1)^{E(x)_i}\ket{1}_i
\eeq
is \textit{exactly} an implementation of the fingerprint state of Eq. \eqref{FPstate}. Here $\ket{1}_i$ denotes a one photon state in the $i$th mode. Since these states are an exact implementation of the fingerprint states, they are equivalent to states of $\order{\log_2 n}$ qubits, even if the number of modes employed is proportional to the input size $n$. This clearly indicates that the amount of abstract communication in a protocol is not given by the number of modes used.

In general, we must quantify the amount of communication by the smallest number of qubits that would be required, in principle, to replicate the performance of the protocol. More precisely, if a quantum communication protocol uses states in a Hilbert space of dimension $d$, this space can be associated to a system of $\order{\log_2 d}$ qubits. Therefore, the amount of communication $C$ in a quantum protocol is generally given by
\beq
C=\log_2[\dim(\Hil)]
\eeq
where $\Hil$ is the smallest Hilbert space containing all the states of the protocol, which may be a significantly smaller subspace of the entire Hilbert space associated to the physical systems. For example, a single photon in the polarization degree of freedom can be used as a qubit, but we require two polarization modes, each representing an infinite-dimensional Hilbert space. Moreover, Holevo's theorem \cite{holevo1973bounds} guarantees that no more than $\log_2 d$ classical bits of information could be transmitted, on average, by a quantum protocol that uses states in a Hilbert space of dimension $d$.

By quantifying communication carefully, we gain a better understanding of the different physical resources that are required to transmit a certain amount of information. For example, the fact that the same amount of information can be transmitted by a single photon in $n$ optical modes, at most $n$ photons in a single mode or $\log_2 n$ qubits, is understood because the smallest Hilbert space containing all possible states in each of the three cases has the same dimension.

In terms of an experimental demonstration, creating states of fixed photon number in a superposition of modes, as proposed in Refs. \cite{QFingerprintingMassar,SWAPHOM}, is an extremely challenging task \cite{gauthier2012quantum}. Instead we opt for an alternative that is readily implementable in practice: a coherent state in the fingerprint mode. This \textit{coherent fingerprint state} can be written as $\ket{\alpha}_x=D_x(\alpha)\ket{0}$, where $D_x(\alpha)=\exp(\alpha a_x^{\dagger}-\alpha^*a_x)$ is the displacement operator and $\alpha$ is a complex number. A straightforward calculation shows that this state can be equivalently expressed as a simple sequence of coherent pulses
\beq\label{coherentfpstates}
\ket{\alpha}_x=\bigotimes_{i=1}^m\ket{(-1)^{E(x)_i}\am}_i,
\eeq
where $|\am\rangle_i$ is a coherent state with amplitude $\am$ in the $i$th mode. Notice that a projection of this state onto the single-photon subspace gives exactly the state of Eq. \eqref{FPFockstate}.

The phase of each individual state in the product depends on the corresponding bit of the codeword. Therefore, to implement the states correctly, Alice and Bob need a common phase reference, which can be established before the start of the protocol or may be available already from other contexts, without giving Alice and Bob access to shared randomness. On the other hand, the referee needs a measurement that allows her to verify whether the relative phases of the incoming pulses are equal or different. A way of achieving this consists of an interferometer in which the individual pulses enter a balanced beam splitter and whenever there is a click in the output detectors, it is unambiguously revealed whether their phases are the same or not \cite{PhysRevA.74.022304}. We call these outcomes ``0" and ``1" respectively, in accordance to the relative parity of the phases. In this way, we have established the basic ingredients for a quantum fingerprinting protocol in an ideal implementation: 
\begin{enumerate}
\item{ Alice and Bob fix a value $c$ for the Justesen code and of $\alpha$ for the coherent fingerprint states.}
\item{They prepare the states $\ket{\alpha}_x,\ket{\alpha}_{x'}$ according to their respective inputs $x,x'$ as in Eq. \eqref{coherentfpstates}}.
\item{They send these states to the referee, who performs an interference measurement on the individual signals using a balanced beam splitter and single-photon detectors. }
\item{The referee concludes that the inputs are different if and only if she observes at least one click in the ``1" detector.}
\end{enumerate} 

\begin{figure}
\includegraphics[width=0.8\columnwidth]{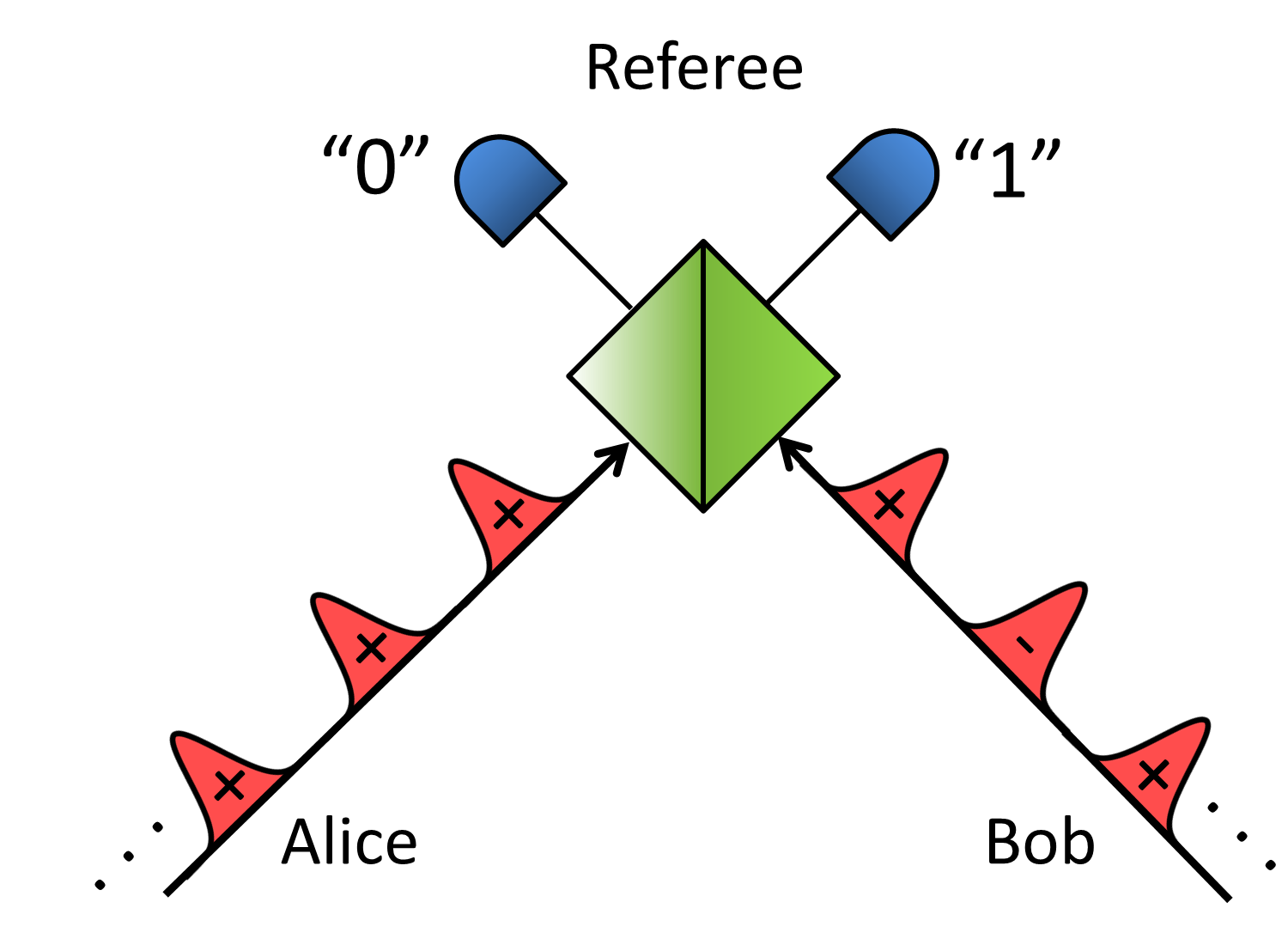}
\caption{(Color online) Coherent-state protocol: Alice and Bob send a train of $m$ coherent pulses whose phases (``+" or ``$-$") depend on the inputs they receive. The referee interferes the individual signals in a 50:50 beam splitter and concludes that the inputs are different if and only if at least one ``1" click is observed.}\label{Interference}
\end{figure}

An illustration of the protocol is shown in Fig. \ref{Interference}. 

As discussed before, the abstract communication cost of a quantum protocol, which is equal to the amount of information transmitted, is determined by the dimension of the quantum states used. In our case, the coherent fingerprint states are effectively contained in a Hilbert space of small dimension, as is formally summarized by the following statement:

\begin{thm}\label{Theorem}
There exist a set of states $\{\ket{v}_x\}$ of dimension $d$ satisfying $\log_2d=\order{\log_2 n}$, such that for any $\epsilon>0$, it holds that $||\ketbra{v}{v}_x-\ketbra{\alpha}{\alpha}_x||_1\leq \epsilon$, for all inputs $x$.
\end{thm}
\textit{Proof:} For a given $\Delta_N$, let $\Hil_V$ be the subspace spanned by the Fock states $\ket{N}_x$ whose photon number $N$ satisfies $|N-|\alpha|^2|\leq\Delta_N$. To calculate the dimension of this subspace, we use the fact that the dimension of the space of states with fixed a photon number $N$ is equal to the the number of distinct ways in which the photons can be distributed into the $m$ different modes. Since the photons are indistinguishable, this quantity is given by the binomial factor $\binom{N+m-1}{m-1}$ \cite{2002first}. In the case of $\Hil_V$, there are $2\Delta_N$ different possible values of $N$, the largest being $N=|\alpha|^2 +\Delta_N$. Thus, the dimension $d$ of this subspace satisfies
\begin{align}
&\log_2 d\leq \log_2\left[2\Delta_N\binom{|\alpha|^2 +\Delta_N+m-1}{m-1}\right]\nonumber\\
&\leq(|\alpha|^2+\Delta_N) \log_2\left(m+|\alpha|^2+\Delta_N-1\right)+\log_2(2\Delta_N),
\end{align}
which is $\order{\log_2 n}$ for any fixed $\alpha$ and $\Delta_N$. 

Now let $P_V$ be the projector unto $\Hil_V$ and define the $\order{\log_2 n}$-qubit states $\ket{v}_x:=P_V\ket{\alpha}_x/||P_V\ket{\alpha}_x||$. Consider a measurement $\{P_V,\id-P_V\}$ that informs us whether the photon number lies within the range $|N-|\alpha|^2|\leq\Delta_N$. Since all of the coherent fingerprint states have the same Poissonian photon number distribution with mean $|\alpha|^2$, we can use the properties of this distribution to calculate the probability that the measured number of photons $N$ deviates by an amount $\Delta_N$ from its expected value. This probability satisfies \cite{franceschetti2007closing}
\begin{align}
&P(|N-|\alpha|^2|\geq\Delta_N)\leq\nonumber\\
&2e^{-|\alpha|^2}\left(\frac{e|\alpha|^2}{|\alpha|^2+\Delta_N}\right)^{|\alpha|^2+\Delta_N}=\epsilon'.
\end{align}
This also implies that $1-|\braket{v}{\alpha}_x|^2\leq \epsilon'$. Finally, using the Fuchs-van de Graaf inequality \cite{fuchs1999cryptographic} we have that
\beq
||\ketbra{v}{v}_x-\ketbra{\alpha}{\alpha}_x||_1\leq 2\sqrt{1-|\braket{v}{\alpha}_x|^2}\leq 2\sqrt{\epsilon'}
\eeq 
and this can be made equal to any $\epsilon>0$ by choosing $\Delta_N$ accordingly while keeping $\alpha$ fixed. \qed

The above result implies that the statistics obtained from any measurement on the coherent fingerprint states can be made arbitrarily close to those obtained from states of $\order{\log_2 n}$ qubits. Therefore, for sufficiently small $\epsilon$, the two cases are operationally indistinguishable and an exponential separation in communication complexity is maintained. 

To calculate the error probability of the protocol, notice that whenever $x=x'$, the referee outputs the correct answer with certainty because the only possible outcomes are ``0" or no clicks. For the case of $x\neq x'$, we need to calculate the probability that no ``1" outcomes are observed. It can be shown (see Appendix for details) that this probability of error satisfies
\beq
P_m(\text{error})\leq [1-p_c(1-\delta)]^m,
\eeq
where $p_c$ is the probability of obtaining a click at each time slot, which is given by
\beq
p_c=1-\exp(-2\amsqr).
\eeq
To illustrate the behaviour of this quantity, note that for large $m$, we can make the approximation $p_c\approx\frac{2|\alpha|^2}{m}$ so that
\begin{align}
P_{m}(\text{error})&\approx \left(1-\frac{2(1-\delta)|\alpha|^2}{m}\right)^m\nonumber\\
&\leq \exp[-2(1-\delta)|\alpha|^2].
\end{align}
Therefore, for fixed $\delta$, the probability of error can be made arbitrarily close to zero by fixing $\alpha$ accordingly, and this error decreases exponentially with $\alpha$. Moreover, we can choose the total mean photon number of the coherent fingerprint states independently of the input size and still satisfy any demand on the error probability.

\textit{Protocol in the presence of experimental errors.-} So far, we have assumed an ideal scenario, but any practical implementation will invariably suffer from the presence of experimental errors. Our goal is now to show that the above protocol can be modified to become robust against these errors.

The main drawback of the previous protocol is that it is extremely sensitive to the error that occurs when the fingerprints are equal, but the ``1" detector fires due to an imperfection. Nevertheless, it is natural to envision a situation in which the expected ratio of ``0" to ``1" clicks differs significantly for the cases of equal or different inputs, so that these situations can be statistically distinguished. Formally, let $f_0$ be the observed fraction of ``0" outcomes and define the expectation values $q_E:=\mathbb{E}(f_0|x=x')$, $q_D:=\mathbb{E}(f_0|x\neq x')$ and $\Delta_q=(q_E-q_D)/2$. The modification to the protocol is then very simple: The referee concludes that the inputs are equal if and only if $f_0>q_E-\Delta_q$. In this case, it can be shown (see Appendix for details) that the probability of error satisfies
\beq
P_m(\text{error})\leq [1-p'_c(1-e^{-2\Delta_q^2})]^m,
\eeq
where $p_c'\approx p_c+p_{dark}$ is the effective click probability. Again, for large $m$, $p'_c\approx 2|\alpha|^2/m$ and we get
\beq\label{perror}
P_m(\text{error})\lesssim e^{-2|\alpha|^2(1-e^{-2\Delta_q^2})},
\eeq
which can also be made arbitrarily close to zero by fixing $\alpha$ accordingly.

The values of the expectations $q_E$ and $q_D$ as well as the click probability $p'_c$ are determined by the experimental errors. We consider a model of imperfections characterized by three parameters: the combined effect of channel loss and limited detector efficiency $\eta$, the limited visibility of the interferometer $\nu$ and the dark count probability $p_{dark}$. The effect of loss and limited efficiency is equivalent to a transformation $\ket{\alpha}_x\rightarrow|\sqrt{\eta}\alpha\rangle_x$, which can always be compensated by increasing the initial value of $\alpha$ to $\alpha/\eta$, without changing the scaling properties of the protocol.

Since $p_{dark}\ll1$ and $p'_c\ll1$ for large $n$, we neglect the occurrence of double-clicks in one time-slot. In this case the expected fractions of ``0" outcomes can be shown to be
\begin{align}
q_D=&\frac{p_c}{p_c+p_{dark}}\left[\nu\delta+(1-\nu)(1-\delta)\right]+\frac{p_{dark}}{2(p_c+p_{dark})}\nonumber\\
q_E=&\frac{p_c}{p_c+p_{dark}}\nu+\frac{p_{dark}}{2(p_c+p_{dark})}.
\end{align}
From these expressions we can also calculate $\Delta_q$ to obtain
\beq
\Delta_q=\frac{p_c}{p_c+p_{dark}}(1-\delta)(2\nu-1).
\eeq
\begin{figure}
\includegraphics[width=\columnwidth]{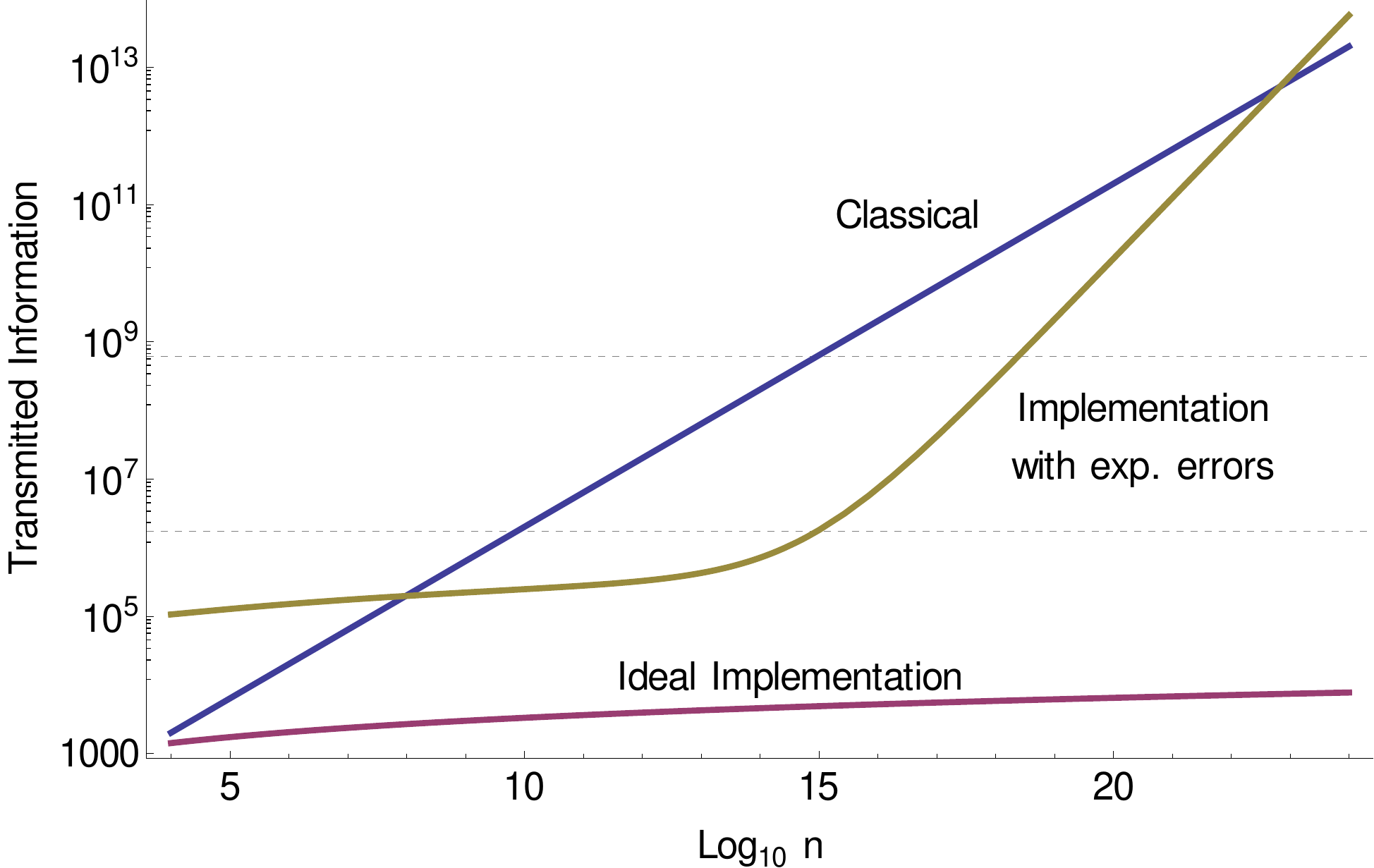}
\caption{(Color online) Logarithmic plot for the transmitted information as a function of the input size $n$ for different fingerprinting protocols with probability of error of $10^{-6}$.  We adopt the classical protocol specified in \cite{babai1997randomized}, which requires $2\sqrt{n}+\order{1}$ bits of communication and must be repeated ten times to ensure the desired probability of error. For the quantum case, we choose $c=3$ for the Justesen code and portray the cost for the coherent-state protocol with an ideal implementation ($|\alpha|^2=88.8$) and for a non-ideal implementation ($|\alpha|^2=6,651$) suffering from experimental errors. The effective dimension of the states is chosen so that the trace distance between the fingerprint states and states of this dimension is smaller than $10^{-6}$. The errors are characterized by the parameters $\eta=0.1$, $\nu=0.98$ as in the experiment of \cite{PhysRevA.68.022317} and by $p_{dark}=4\times 10^{-8}$ as would occur with the SNSPD detectors of Ref. \cite{marsili2013detecting}. For $n\sim 10^{13}$, it is not possible to maintain the desired error probability for fixed $\alpha$, and increasing the mean photon number leads to a steeper increase in the transmitted information. }\label{Plots}
\end{figure}
It is important to notice the crucial role played by the dark count probability, which sets a limit on the maximum input size the protocol can tolerate with a fixed mean photon number. When the click probability $p_c$ becomes smaller than $p_{dark}$, most of the outcomes are random regardless of whether the inputs are equal or different, making the two situations increasingly difficult to distinguish. To put this into context, it is currently possible to achieve values as low as $p_{dark}\sim 10^{-8}$ \cite{marsili2013detecting}. In this case, the protocol can function for input sizes of up to $n\sim 10^{13}$ for a constant mean number of photons. This can be seen in Fig. \ref{Plots}, were a comparison of classical and quantum fingerprinting protocols is made. We also highlight that even in the presence of errors, our protocol surpasses the performance of a classical protocol \cite{babai1997randomized} for a wide range of input sizes, with a reduction in the transmitted information that can be larger than two orders of magnitude, as depicted by the dotted lines in Fig. \ref{Plots}.

\textit{Discussion.-} 
We have outlined a quantum fingerprinting protocol that can be implemented with current technology \cite{rubenok2013real}, even in the presence of experimental imperfections and demonstrates an exponential separation in communication complexity compared to the classical case. Previous work had proposed different paths towards the implementation of quantum fingerprinting, but none of them could be experimentally deployed to the point of exhibiting a gap in communication complexity compared to the classical case. 

From a practical perspective, we are often interested in the expenditure of resources beyond the abstract amount of communication. For instance, we may be interested in the running time of the protocol or the amount of energy used. Since our protocol uses $\order{n}$ optical modes, the total time required to carry the protocol is quadratically larger than what would be needed in a classical protocol. On the other hand, the total number of photons used is constant, whereas classically one would need $\order{\sqrt n}$ photons when restricted to use $\order{\sqrt n}$ modes. Thus, our protocol introduces an asymptotically unbounded reduction in energy consumption for the price of only a quadratic increase in running time. Moreover, by Theorem \ref{Theorem}, any classical communication protocol using $\order{n}$ modes and a constant photon number could only be used to transmit $\order{\log_2 n}$ classical bits of information, which is insufficient to solve the equality problem in the simultaneous message passing model. This also means that only $\order{\log_2 n}$ classical bits of the input bit strings are leaked to the referee (or anyone else). Overall, given the expended resources, our protocol achieves a task that is provably impossible with classical communication only.

The fact that the total mean photon number is constant has potential practical implications beyond the inherently vast reduction in energy consumption. The clock rate of a quantum communication protocol is usually limited by the dead times of the detectors. However, since in our case each individual mode carries very few photons on average, the expected time between detector clicks could be significantly larger than the dead times, allowing an increase of the clock rate by orders of magnitude. Moreover, time resolution is unnecessary in our scheme, only the click patterns matter regardless of the times at which they occur. Finally, the low photon numbers imply that nonlinear effects are not an issue in the transmission of the signals.

Most importantly, the fact that only a small subspace is employed in our scheme implies that, in principle, the unused sections of the entire Hilbert space can still be used for other purposes such as the transmission of additional information through multiplexing schemes. For example, it may be possible to conduct multiple quantum fingerprinting protocols in parallel or perform them alongside classical communication. Although this multiplexing can be achieved in principle, practical methods for achieving it are a line for future research.

Generally, our results imply that any state $\ket{\psi}=\sum_{i=1}^d c_i\ket{i}$ can be approximately implemented by a sequence of coherent states $\otimes_{i=1}^d \ket{\alpha c_i}$. This can provide a promising route for the implementation of other quantum communication protocols. An example are existing schemes for quantum digital signatures \cite{clarke2012experimental,dunjko2014quantum} that also use sequences of phase-encoded coherent states. Our fingerprinting protocol also provides a new ground in which to explore fundamental aspects of quantum mechanics, such as the connection between entanglement and non-orthogonality, the information-carrying capacity of quantum states and the regime of extremely low mean photon numbers. Overall, our results pave the way for experimental demonstrations of the gap between classical and quantum communication complexity, and open a new window of opportunity for research in quantum communication in general.

We thank R. Cleve, R. Koenig, M. Curty, H. Buhrman and W. Tittel for fruitful discussions concerning this work. J.M. Arrazola would like to thank A. Ignjatovic for her help in preparing this manuscript and he acknowledges support from the Mike and Ophelia Lazaridis Fellowship. This work has been supported by Industry Canada, the NSERC Strategic Project Grant (SPG) FREQUENCY and the NSCERC Discovery Program.

\appendix*
\section*{Appendix}
\textit{Error probability for the ideal protocol.-} To calculate the error probability of the ideal protocol, notice that whenever the inputs to Alice and Bob satisfy $x=x'$, the referee outputs the correct answer with certainty because the only possible outcomes are ``0" or no clicks. For the case of $x\neq x'$, we need to calculate the probability that no ``1" outcomes are observed. After the individual signals interfere in the beam splitter, there will always be a coherent state entering one detector and the vacuum entering the other one. The probability $p_c$ of obtaining a click can be calculated from the Poissonian statistics of the incoming coherent states and is given by
\beq
p_c=1-\exp(-2\amsqr).
\eeq where $p_c\approx 2\frac{|\alpha^2|}{m}$ for $\frac{|\alpha^2|}{m}\ll 1$. The total number of clicks in the $m$ signals is therefore a binomial random variable, which we call $C$. We introduce another random variable $Z$, the number of ``0" outcomes observed. When $x\neq x'$, the conditional probability distribution of $Z$ given that $k$ clicks are observed is a hypergeometric distribution satisfying
\beq
P_{m}(Z=\ell|C=k) =\frac{\binom{m\,\delta}{\ell}\binom{m-m\,\delta}{k-\ell}}{\binom{m}{k}}.
\eeq 
In this case, the probability of error is given by
\begin{align*}
P_{m}(\text{error})&=\sum_{k=0}^m P(C=k)P(Z=k|C=k)\\
&=\sum_{k=0}^{m} \binom{m}{k}p_c^k(1-p_c)^{m-k}\frac{\binom{m\,\delta}{\ell}\binom{m-m\,\delta}{k-\ell}}{\binom{m}{k}}\\
&\leq \sum_{k=0}^{m} \binom{m}{k}(p_c\delta)^k(1-p_c)^{m-k},
\end{align*}
where we have used the inequality 
\beq
\frac{\binom{m\,\delta}{\ell}\binom{m-m\,\delta}{k-\ell}}{\binom{m}{k}}\leq\delta^k,
\eeq
which can be proven with a straightforward calculation. From the binomial theorem we conclude that for any two inputs
\beq
P(\text{error})_{m}\leq [1-p_c(1-\delta)]^m.
\eeq

\textit{Error probability for the protocol in the presence of experimental errors.-} To bound the probability of error in this case, we consider first the case $x=x'$ and denote by $p'_c\approx p_c+p_{dark}$ the effective click probability. We then have
\begin{align*}
&\Pr(f_0\leq q_E-\Delta_q|x=x')\\
=&\sum_{k=0}^{m}P(C=k)P(f_0\leq q_E-\Delta_q|k,x=x')\\
=&\sum_{k=0}^{m}\binom{m}{k}(p'_c)^k(1-p'_c)^{m-k}\Pr\left(\frac{Z}{k}\leq q_E-\Delta_q|x=x'\right)\\
\leq&\sum_{k=0}^{m}\binom{m}{k}\left(e^{-2 \Delta_q^2}p'_c\right)^k(1-p'_c)^{m-k}\\
=&[1-p'_c(1-e^{-2\Delta_q^2})]^m,
\end{align*}
where we have made use of Hoeffding's inequality \cite{hoeffding63a}
\begin{align*}
\Pr\left(\frac{Z}{k}\leq q_E-\Delta_q|x=x'\right)&\leq e^{-2k \Delta_q^2},
\end{align*}
which holds when $Z$ is hypergeometrically distributed. Since the Hoeffding bound for $\Pr(f_0\leq q_E-\Delta_q|x=x')$ is equal to that of $\Pr(f_0\geq q_D+\Delta_q|x\neq x')$, the bound on the probability of error is the same when the fingerprints are different. Therefore we can conclude that
\beq
P_m(\text{error})\leq [1-p'_c(1-e^{-2\Delta_q^2})]^m.
\eeq
\bibliography{Refs}
\bibliographystyle{apsrev}

%%%%%%%%%%%%%%%%%%%%%%%%%%%%%%%%%%%%%%%%%%%%%%%%%%%%%%%%
\end{document}